\documentclass{ws-ijmpcs}
\usepackage{cite}
\begin{document}

\markboth{Jean-Marc Richard}
{Baryon spectroscopy and heavy quarks}

%
\catchline{}{}{}{}{}
%

\title{Baryon spectroscopy and heavy quarks\footnote{Invited talk at ``Charm 2010'', 4th Int. Workshop on Charm Physics, Oct.\  21-24, 2010, IHEP, Beijing, China, to appear in the Proceedings, ed.\ Hai-Bo Li et al., to be published in IJMPCS}
}

\author{Jean-Marc RICHARD}

\address{Universit\'e de Lyon and Institut de Physique Nucl\'eaire de Lyon, IN2P3-CNRS--UCBL\\
              4, rue Enrico Fermi, F-69622 Villeurbanne, France\\
j-m.richard@ipnl.in2p3.fr}

\maketitle

\begin{history}
\received{Day Month Year}
\revised{Day Month Year}
\end{history}

\begin{abstract}
Two topics are briefly reviewed in this talk: the decay of flavored hadrons or quarkonium states involving a baryon--antibaryon pair, and the spectroscopy of heavy baryons containing one, two or three heavy quarks. Some prospects for exotic heavy baryons are also discussed.

\end{abstract}

\ccode{PACS numbers: 11.25.Hf, 123.1K}

\section{Introduction}	

Heavy quarks and baryons have a long and rich joint history. Even before the discovery of charm (but published shortly after), a comprehensive article by Gaillard et al.\cite{Gaillard:1974mw}\ anticipated the rich baryon spectroscopy  when the charmed quark is used as a new constituent.  The experimental knowledge of baryons with a single heavy quark started rather early, with the discovery of the $\Lambda_c$ in 1976, and has made much progress in recent years. Unfortunately, there are uncertainties about double-charm baryons, and there is not much emulation to look at this sector, though it should be accessible with existing beams and detectors.

The decay of heavy quarkonia and heavy flavored mesons involving a baryon and an antibaryon gives useful information on the strong and weak annihilation mechanisms, and on the spectroscopy of light baryons in a limited mass range. This will be discussed first.

\section{Light baryons from heavy quark decays}

In 1980, Pham proposed the reaction $D_s\to p+\bar n$ as a probe of the annihilation mechanism in weak decays\cite{Pham:1980dc}\@ This rare mode has been discovered recently.\cite{Athar:2008ug}

The $J/\psi$ decay into a proton and an antiproton has been identified and measured shortly after the discovery of charmonium. See, e.g.,Ref.~\refcite{Nakamura:2010zzi} for references. The proton--antiproton coupling  has been used to refine charmonium spectroscopy at CERN, in the last experiment hosted by the  ISR, and at Fermilab. It will be used in the planed PANDA experiment.\cite{Gianotti:2010zz}

Meanwhile, the charmonium decay measurements have been extended as to include other baryon--antibaryon pairs, and $\psi(2S)$ decay. Once phase-space factors are removed, the branching ratios for $p\bar p$, $n\bar n$, \dots, $\Xi\bar\Xi$, and even $\Delta\bar\Delta$ are nearly identical, and demonstrate an almost perfect SU(3) symmetry, and even a kind of SU(6) symmetry. This should be kept in mind when discussing the so-called ``strangeness suppression'' effects in low-energy hadronic processes. 

\section{Weak decay of heavy baryons}

This was a shock in our community when the ratio $R$ of $D^+$ to $D^0$ lifetime was announced to be $R\simeq 4$! Even the revised and stabilized value $R\simeq 2.5$\cite{Nakamura:2010zzi} clearly contradicts the current wisdom of the 70s that the charmed quark, while decaying, ignores its environment. It is now understood that, besides the spectator diagrams, there are $W$-exchange diagrams for $(c\bar d)$ and $(c\bar s)$, and interference effects between  the $\bar d$ ($\bar s$) antiquark  produced in $c\to s+W^+$ and $W^+\to u+ \bar d\ (\bar s)$ in $(c\bar d)$ decay, to a lesser extent, for the Cabibbo-suppressed decay of $(c\bar s)$. As already mentioned, there is an annihilation diagram for $(c\bar s)$.

\begin{figure}[htp]
\centerline{\psfig{file=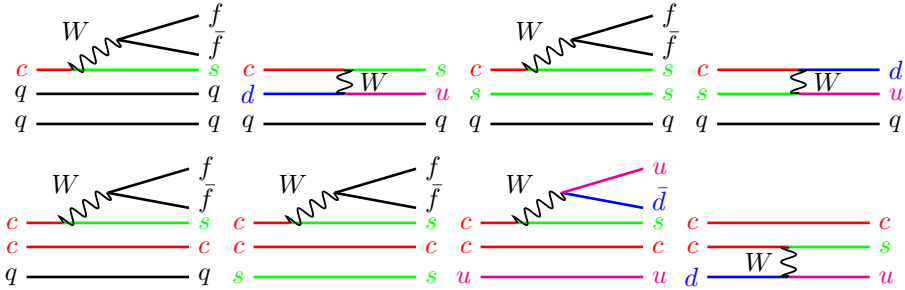}
}
\caption{Some contributions to the weak decay of singly and doubly-charmed baryons: spectator,  spectator with interference,
$W$-exchange in the main or in the Cabibbo-suppressed sector,  and $W$-exchange .}
 \label{fig:bar-dec}
\end{figure}

The same mechanisms have been applied to the decay of baryons. See, e.g., Ref.~\refcite{Chang:2007xa} for one of the latest contributions and references there to earlier works. The annihilation diagram is of course suppressed. There is another type of interference, between the $s$ quark originating from $c$ decay (or $d$ for Cabibbo-suppressed transitions) and a constituent $s$ (or $d$) quark from the decaying baryon.  See Figs.~\ref{fig:bar-dec}.

A hierarchy of lifetimes has been predicted for the various single-charm baryons and also a hierarchy of hadronic-to-leptonic branching ratios. The measurements are in good agreement with the predictions, except that the observed differences are even more pronunced. Altogether there is more than one order of magnitude between the shortest ($\Omega_c$) and longest ($D^\pm$)  lifetime of single charm!

For double charm, there is now some consensus\cite{Chang:2007xa}  that
\begin{equation}
 \tau(ccs]\lesssim\tau(ccd)<\tau(ccu)~,
\end{equation}
where, as for $(ccc)$ and for $(b\bar c)$ in the meson sector, binding effects cannot be neglected. 

The precise ordering of $\tau(ccs)$ vs.\ $\tau(ccd)$ is more sensitive to the details. As for the semi-leptonic branching ratios, the expectation is\cite{Chang:2007xa,Guberina:1999mx}
\begin{equation}
R(ccd)<R(ccs)<R(ccu)~.
\end{equation}

\section{Spectroscopy of heavy baryons}
\paragraph{Single heavy flavor}
Several new results came in recent years from $e^+e^-$ colliders and from Fermilab.  With the flavor-independent confinement, one expects a strict similarity between the $(cqq)$ and $(bqq)$ spectra. In the potential models, this comes from the masses being dominated by the light quarks. This is the same mechanism that makes ${}^3\mathrm{He}$ and ${}^4\mathrm{He}$ electronic spectra nearly identical in atomic physics. In more elaborate frameworks, this property is one of the aspects of the heavy quark symmetry, and the systematics of $1/M_Q$ correction can be written down.\cite{Albertus:2005uq}

Somme differences between $(cqq)$ and $(bqq)$ were anticipated in constituent models, and recovered, e.g., by QCD sum  rules.\cite{Albuquerque:2009pr}\@ In particular, while the $\Sigma_Q-\Lambda_Q$ mass difference does not change much as $M_Q$ increases, due to the chromomagnetic interaction among light quarks, the $\Sigma_Q^\star-\Sigma_Q$ is expected to decrease. 

If a comparison is made of the lowest flavor and spin excitation of single-charm and single-beauty baryons, the splittings are found in good agreement. However, the $\Omega_b$ candidate by D\O\ is substantially higher than anticipated from the mass difference $\Omega_c-\Lambda_c$, while the CDF candidate  mass looks more plausible. 
See, e.g., Refs.~\refcite{Albuquerque:2009pr,Klempt:2009pi} and references there.

There are many interesting features in the excitation spectrum. For instance, while the lowest excited $\Xi_c(csq)$ states are seen to decay to a lower $\Xi_c$ and a pion, the higher states use the modes with $\Lambda_c$ and kaon,\cite{Aubert:2007dt} to be tentatively interpreted in terms of the chromo-electric field and probability of $q\bar q$ pair-creation in the $c-s$ vs.\ $(cs)-q$ strings.\cite{Eakins:2010zz}

\paragraph{Double heavy flavor}

The  experimental situation is rather embarrassing.  Selex  has published evidence for the $\Xi_{cc}^+$, from its weak decay, either with the remaining charm in a baryon,\cite{Mattson:2002vu} or in a meson,\cite{Ocherashvili:2004hi} but this has not been confirmed elsewhere. In particular, Babar\cite{Aubert:2006qw} and Belle\cite{Chistov:2006zj}  searched unsuccessfully for double-charm baryons. The situation is not \textsl{a priori} favorable  in $e^+e^-$, but since the double-charm production has been observed under the form of $(c·\bar c)+(c\bar c)$, it could be naively expected that the same mechanism sometimes leads to $cc+cc$, and eventually to double-charm baryons after hadronization. 
As for  hadron beams, an experiment was proposed fourteen years ago, and approved with  ``charmed hadrons'' and in particular,  ``double charm baryons'' among the top priorities, but  has not yet  touched this field. Meanwhile, many discoveries have been made elsewhere in heavy-quark spectroscopy, and many others are awaiting for a serious investigation with hadron beams. 

The double-charm baryons are probably the most interesting of ordinary hadrons, as they combine two extreme regimes of QCD: the adiabatic motion of two heavy quarks, and the relativistic motion of a light quark around a colored source.  There  is now an abundant literature on $(QQq)$ baryons, including QCD sum rules\cite{Narison:2010py,Wang:2010vn} and 
lattice QCD.\cite{Lin:2010wb}\@
For the constituent-model approach, see, e.g., Ref.~\refcite{Vijande:2004at}. Note that a quark--diquark approximation is tempting, as the two heavy quarks tend to cluster in $(QQq)$. However, as the lowest excitations lie mainly within $QQ$,  a new diquark is required for each level. In contrast, the Born--Oppenheimer approximation is more efficient, as for $\mathrm{H}_2^{\ +}$ in atomic physics. This was shown within a simple potential model,\cite{Fleck:1989mb} but one can estimate the effective $QQ$ potential from the lattice.

\paragraph{Triple heavy flavor}

\begin{figure}[htp]
\hspace*{.25cm}
\begin{minipage}{.65\textwidth}
\psfig{file=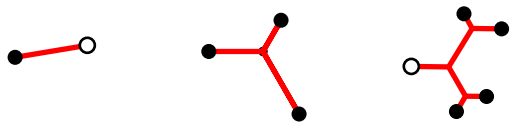}
\hspace*{.5cm}
\end{minipage}
\
\begin{minipage}{.28\textwidth}
 \caption{String limit of the confining interaction for mesons, baryons, and penta\-quarks.}
 \label{fig:string}
\end{minipage}
\vspace*{-.8cm}
\end{figure}

This is a challenging goal for experimentalists, as the final multiplicity is rather large, to fully reconstruct $(ccc)$, $(ccb)$ or $(bbb)$. These configurations have been estimated first with potential models, and now on lattice.\cite{Meinel:2010pw}\@ One expects that the first excitation should be of negative parity, unlike the case of light baryons, for which the Roper resonance, with positive parity, occurs first. The $(QQQ)$ potential should be the analog of the famous $V=-a/r+b\,r$ quarkonium potential.  A part  of the Coulomb term presumably comes from one-gluon exchange, and should be extended as $\sum 1/r_{ij}$ for the baryon case, with a color factor $1/2$ in front of its strength. There is now some agreement that this ``1/2'' rule does not hold for the linear part, though it might be a reasonable approximation, the proper generalization of linear confinement being a three-body interaction with $Y$ shape, a string linking the three quarks with minimal cumulated length, as sketched in Fig.~\ref{fig:string}.  As stressed in other contributions to this Workshop, the generalization of the $Y$ shape to multiquark configurations leads to dramatic predictions of  stable exotics. 
%
\paragraph{Exotic heavy baryons}
There has been several proposals to build exotic baryons. Hybrid baryons, for instance, are on the same footing as hybrid mesons, but they do not bear exotic quantum numbers. 

The possibility of baryons with strangeness $S=+1$ was suggested in the 60s, and more recently, for the light pentaquark. 

In 1987, a $(\bar Q qqqq)$ configurations, with $qqqq$ in a spin 0 and flavor-SU(3) triplet was suggested, as a consequence of coherences in the chromomagnetic interaction.

References can be found in Ref.~\refcite{Richard:2009rp}, where another possibility of stable multiquark baryons is envisaged, based on an extension of the string model, see Fig.~\ref{fig:string}. So far, the model has been applied only to quarks of different flavors, to avoid the difficulties due to antisymmetrization.  The pentaquark is found stable in a large range of mass ratios.  The case of identical quarks is currently under study. 
%
%
\section*{Acknowledgments}

It is a pleasure to thank our hosts for this very stimulating conference and the warm discussions in a cold temperature, in particular with S.~Brodsky on the issue of double-charm baryons, and also M.~Asghar for comments on the manuscript.
%

\end{document}